\documentclass[12pt]{iopart}

\usepackage{iopams}
\usepackage{mathrsfs}
\usepackage{graphicx}

\begin{document}


\title{Organization mechanism and counting algorithm on Vertex-Cover solutions}


\author{Wei Wei, Renquan Zhang$^{*}$, Baolong Niu, Binghui Guo, Zhiming Zheng}
\address{LMIB and School of mathematics and systems science, Beihang University, 100191, Beijing, China}

\ead{$^{*}$zhangrenquan09@smss.buaa.edu.cn}


\date{\today}

\begin{abstract}
Counting the solution number of combinational optimization problems
is an important topic in the study of computational complexity, especially
on the \#P-complete complexity class. In this paper,
we first investigate some organizations of Vertex-Cover unfrozen subgraphs
by the underlying connectivity and connected components of unfrozen vertices.
Then, a Vertex-Cover Solution Number Counting
Algorithm is proposed and its complexity analysis is provided, the results of which
fit very well with the simulations and have better performance than those by 1-RSB
in a neighborhood of $c=e$ for random graphs.
Base on the algorithm, variation and fluctuation on the solution number
statistics are studied to reveal the evolution mechanism of the solution numbers.
Besides, marginal probability distributions on the solution space
are investigated on both random graph and scale-free graph
to illustrate different evolution characteristics of their solution spaces.
Thus, doing solution number counting based on graph expression of solution space should be
an alternative and meaningful way to study the hardness of NP-complete and \#P-complete
problems, and appropriate algorithm design can help to
achieve better approximations of solving combinational optimization problems and the corresponding counting problems.

\end{abstract}


\maketitle

\section{Introduction}
Vertex-Cover problem is one of the six basic NP-complete problems \cite{cook,garey}
and has a large range of applications such as immunization strategies in
networks \cite{immu}, the prevention of denial-of-service attacks \cite{denial} and monitoring
of internet traffic \cite{moni}. To solve Vertex-Cover
instances efficiently and have better understanding of its typical solution
space (ground states) structures, is considered as a kernel way to probe
the essence of computational complexity, which is highly concerned by many mathematicians,
physicists and computer scientists.

Till now, a large number of algorithmic and theoretical results have been obtained, to investigate
the ratios of minimal vertex-covers for random graphs and how to solve the Vertex-Cover instances efficiently \cite{vc-bp}.
One of the important results is the complexity phase transition for solving Vertex-Cover instances, that is,
a random graph instance can be easily minimally covered by a leaf-removal algorithm \cite{leafremoval}
with high probability when its average degree $c<e$ and the algorithm fails when $c>e$ with high probability.
The complexity phase transition on Vertex-Cover is strongly correlated
with the replica symmetry breaking phenomenon \cite{martin1,martin2}, and when $c>e$ the ground states
collapse into many different clusters. The evolution of the ground-state structures of Vertex-Cover
is assumed to undergo replica symmetry and further-step replica symmetric breaking phases \cite{wolf},
which greatly differs from that of 3-SAT \cite{mezard1}. In 3-SAT problem, the core difficulty is assumed to be
the clustering phenomenon \cite{sat-clustering} and the backbone structure \cite{sat-backbone}. But in Vertex-Cover, at least the backbone structure
is not the key difficulty for solving as it appears for simple instances, and the clustering of minimal vertex-covers
has an obscure organization which is unknown but at least more complicated than the one-step replica symmetric breaking \cite{wolf}.
However, the organization of
ground states in further-step replica symmetric breaking phases is far from being
clearly understood.

Beyond solving the NP-complete problems, another interesting problem related to
NP-class problems is the counting problem, which counts the number of optimizations/solutions of NP-class problems.
The counting problem belongs to an important complexity class in the research of computational complexity, and has profound significance
in investigating the relationship of P and NP problems \cite{sharp-p}.
As one statistical characteristic of the solution space of Constraint Satisfaction Problems (CSPs),
the solution number calculating which is known as \#CSP \cite{n-csp} and corresponds to the \emph{entropy} in statistical physics, should be strongly correlated with the solution space structures. \#CSP problems,
such as \#SAT and \#Graph Colorings \cite{n-sat,n-color}, belong to the \#P complexity class,
and solving the \#Vertex-Cover is \#P-complete \cite{sharp-p} which is at least as hard as the NP-complete class.
Evidently, counting all the answers of CSP problem is quite a difficult job, even for 2-SAT.
The methods of cavity and mean field can be used to calculate the entropy of the ground-state space \cite{vc-entropy},
which has direct correspondence with the counting problems but is still a statistical and approximated one.

In statistical mechanics, the relationship between solution space structures and entropy has been investigated \cite{vc-entropy,space-vc},
and especially the entropy calculations under the assumptions of replica symmetry and one-step replica symmetric breaking
have been performed on some classical problems such as 3-SAT and Vertex-Cover \cite{vc-bp,mezard1}. However, these researches
are generally fulfilled by statistics under assumptions, and till now most strictly proved results strongly rely on tree structures or relatively simple graphs,
which provide little insight on the exact number counting of the solutions.
By the results in \cite{space-vc}, a description
on the solution space, named reduced solution graph for Vertex-Cover, is proposed.
The reduced solution graph $S(G)$ based on the given graph $G$ can provide
a detailed expression on the status of each vertex in the solution space, that is,
the covered backbones, the uncovered backbones and the unfrozen vertices with their connections.
And, when the given graph $G$ has no leaf-removal core, the reduced solution graph $S(G)$ can
exactly express the solution space. Based on this fact, we do some further analysis on
the solution space of Vertex-Cover problem to see the organization of the solutions, build connections
between the reduced solution graph and solution number counting, and investigate the solution number
statistics such as fluctuations and marginal probabilities in this paper.

\section{Statistical analysis of the solution organization}

As an important solution space structure of Vertex-Cover defined in \cite{space-vc}, mutual-determination
reveals the relation of two unfrozen vertices which can determine each other mutually,
and it can help to achieve the reduced solution
graph by a named Mutual-determination and Backbone Evolution Algorithm (MBEA).
Here, the exactness for reduced solution
graph does not rely on the assumption of replica symmetry or replica symmetric breaking, and even if
a graph has many small local cycles but no leaf-removal core, the exactness can also be guaranteed.
In this section, some statistical characteristics of the
underlying solution space will be discussed based on the reduced solution graph.


\subsection{Structural statistics of the unfrozen vertices}
As the backbones on the graph make little contribution to the relationship among the vertices,
what should be concerned is the \emph{unfrozen subgraph} (the unfrozen
vertices with their connections) without frozen vertices on $S(G)$, on which the double edges are used to denote the mutual-determination relations and
single edges are retained to connect unfrozen vertices on $S(G)$. Based on the leaf-removal process,
we can define the leaf-removal levels for the unfrozen subgraph, shown in Fig.1. Evidently, the vertices in the top level can
produce great influence on those in the lowest level.

\begin{figure}[!t]
\centering
\includegraphics[width=5in]{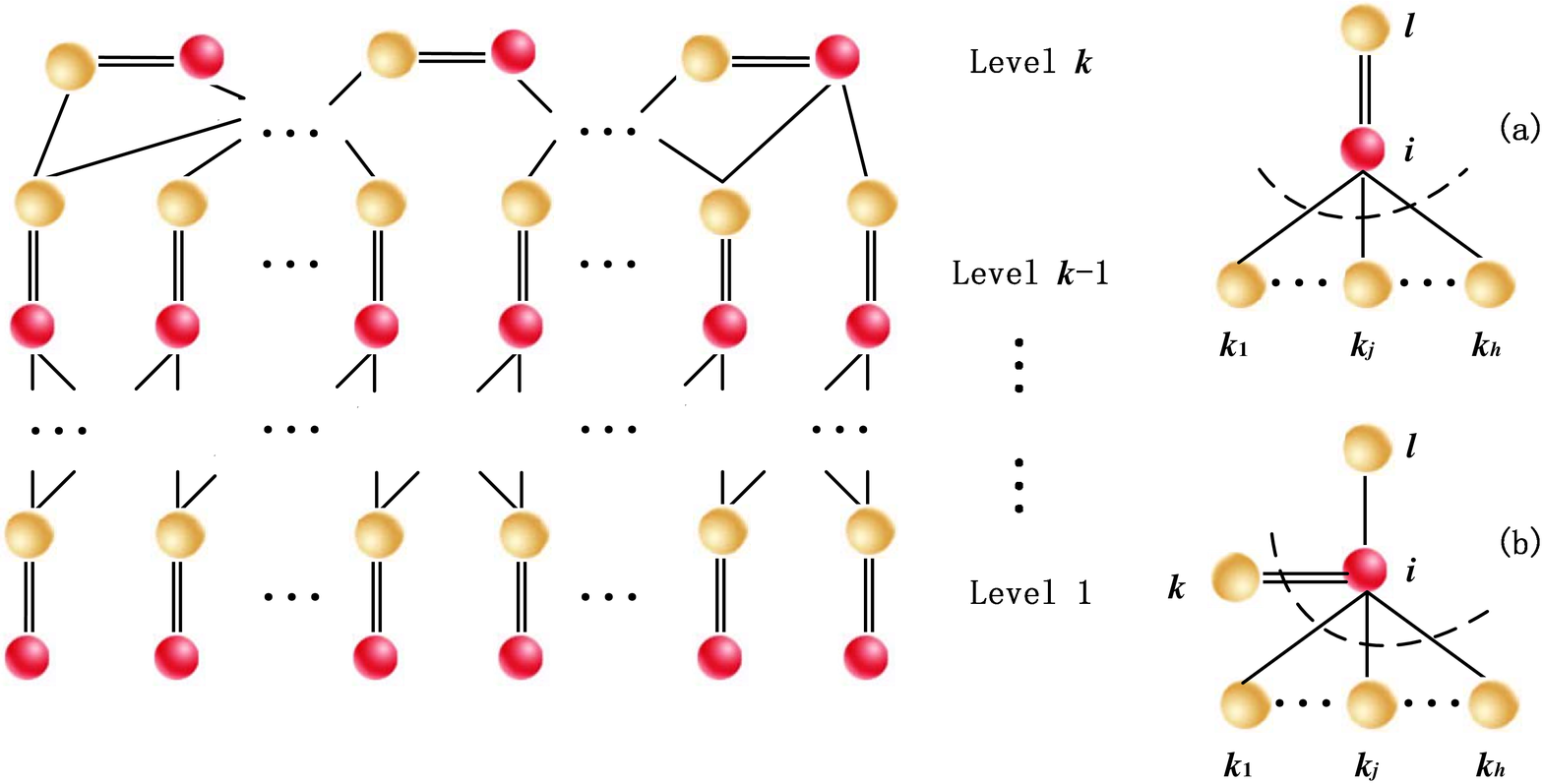}
\caption{Left figure: leaf-removal levels for the unfrozen subgraph. In each level, there are
some mutual-determinations, and connections between different levels are retained by original
single edges of $S(G)$. Red vertices in higher levels will become new leaves when all mutual-determinations in the lower levels are removed. Right
figure: the local evolution of the solution number calculation.}
\label{fig1}
\end{figure}

To see topological structures of unfrozen subgraphs, we take advantage of the mean field and cavity methods
to understand the distribution and organization of mutual-determinations.
For convenience, we use symbols $q_0, q_+,q_-$ to represent the ratios of unfrozen vertices,
positive backbones and negative backbones on the reduced solution graph, and evidently
$q_0+q_++q_-=1$.
Using the analysis and by a vertex-adding process in \cite{space-vc,long-range-zhou},
a new added vertex should be unfrozen and can connect other $k$ unfrozen ones only when it has one positive and $k-1$
unfrozen neighbors with the rest neighbors being negative backbones, so its probability for random graphs can be obtained by
\begin{equation}
F_r(k)=\sum_{i=k}^\infty P_r(i)\cdot C_i^1 \cdot q_+ \cdot C_{i-1}^{k-1}\cdot q_0^{k-1} \cdot q_-^{i-k},
\end{equation}
where $P_r(i)=e^{-c}c^i/i!$ is the degree distribution of random graphs.
Here, we neglect the correlations among the neighbors in obtaining equation (1).
In the insets of Fig.2, theoretical results of $F_r(k)$ with numerical ones are provided
to show the validity of equation (1).
The ratio of free edges $q_{edg}$ (double and single edges connecting unfrozen vertices)
can be obtained by the average $q_{edg}=\sum_{k=1}^\infty k\cdot F_r(k)/2$, and the comparison of $q_{edg}$ with unfrozen vertices $q_0$ is given in Fig.2,
which shows that the number of free edges increases over unfrozen vertices at $c=e$.

For the organization of unfrozen subgraph, the double edges connect vertices of mutual-determinations and
single edges connect different mutual-determinations. Then, it is indicated that $q_{edg}$ free edges involve $q_0/2$ double edges
and $q_{edg}-\frac{q_0}{2}$ single edges.
Thus, for each unfrozen vertex, there are other $2\cdot (q_{edg}-\frac{q_0}{2})/q_0$ unfrozen neighbors
in average except the mutual-determination neighbor.
For $c<e$, the unfrozen subgraph must have almost tree structure to
avoid giant component, otherwise long-rang correlations should exist \cite{long-range-zhou} and it will conflict with the effectiveness of replica
symmetry assumption \cite{martin1}.
At $c\geq e$, we have $q_{edg}\geq q_0$, which implies the unfrozen subgraph can not keep the tree structures and a large
quantity of cycles emerge. Thus, by the theory of random graph \cite{rg}, there must be some giant component on the unfrozen subgraph which
has local tree-like structure. This phenomenon reveals that the
emergence of long-range correlations \cite{long-range-zhou,long-range-pnas} in Vertex-Cover is due to the formation
of unfrozen giant component, and by the increase of
$q_{edg}-\frac{q_0}{2}$ single edges among mutual-determinations, the unfrozen subgraph gets more closely connected and involves more cycles.

\begin{figure}[!t]
\centering
\includegraphics[width=5in]{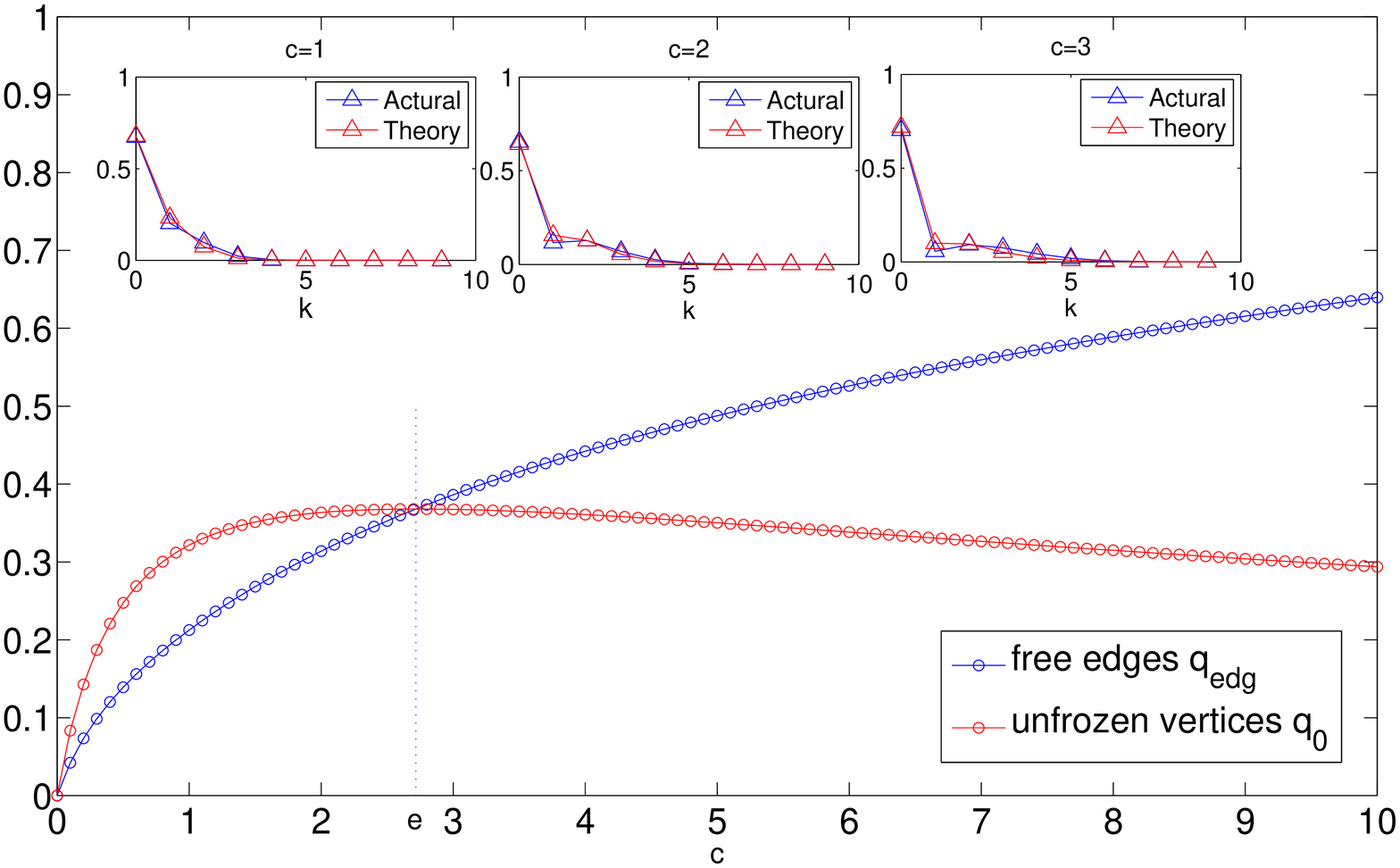}
\caption{The ratios of unfrozen vertices and free edges versus the average degree $c$.
 The insets provide the theoretical and numerical results of degree distributions of the unfrozen vertices for $c=1,2,3$ respectively.
 The results are averaged by 10000 instances with $n=5000$ vertices.}
\label{fig1}
\end{figure}

By the above results on the number of free edges and unfrozen vertices, as the unfrozen
subgraph must be almost tree structures when
$c<e$ for random graphs, almost all the connected components should also be
of tree structure and their number $N_{compo}$ should nearly be $N_{compo}=q_0-q_{edg}$.
Thus, we can deduce that quite a large number of connected components on the unfrozen subgraph are
connected to form one giant component when the average degree reaches $c$.
Similar as random graphs \cite{rg}, on the unfrozen subgraph it can be suggested that small
unfrozen components are absorbed by larger components (especially the maximal ones) as $c$ increases, and the sizes of
maximal unfrozen components can gradually grow when $c<e$ and possess a dramatic increase in the region of $c=e$.

\subsection{Macro- and micro-organizations on the reduced solution graph}

In this subsection, the evolution of giant component as macro-organization and some discussions on vertex status environment
as micro-organization will be studied on the unfrozen subgraph or reduced solution graph.

Here, the evolution of the connected components organizations on unfrozen subgraph is quite different from that of
random graphs. On the random graphs, small connected components gradually increase in random modes and different ones
are connected by a probability determined by their sizes \cite{compo-size}; but on the unfrozen subgraph, the increases of small
connected components slow down at the greatest extent, and the emergence of giant connected component is postponed from $c=1$
for random graph connectivity to $c=e$ for the unfrozen vertices connectivity. Thus, for the evolution modes of the underlying solution space
structure, there should be quite different mechanisms on how the connected components grow. As we know, many component-formation
mechanisms are studied for understanding the generic percolation processes such as the Bohman-Frieze-Wormald (BFW) model \cite{bfw}
and the Achlioptas process \cite{achlioptas}, in which the evolution rules of connected components are mainly constructive and with human interventions.
As an interesting natural underlying graph model, research on the unfrozen subgraph may provide heuristic
ways for the study of the generic percolation processes.

The giant component on the unfrozen subgraph emerges at $c=e$,
which accords with the easily-solving phase transition point \cite{martin2}. By the viewpoint of leaf-removal,
the leaf-removal core exists only after the emergence of the unfrozen giant component, and all the status of removed
leaves can be easily determined by the MBEA algorithm in \cite{space-vc}. For the vertices in the leaf-removal core,
\emph{proper selection} of some covered backbones combined with the mutual-determinations can lead to a consistent and compatible
organization of the frozen and unfrozen vertices,
which can also be expressed as a reduced solution graph. When $c\geq e$, there can be different selections of the
covered backbones in the leaf-removal core,
and each effective selection corresponds to a reduced solution graph and the expression of a solution sub-space.
As unfrozen giant component exists in one such expression, long-range correlations also work, which implies that the replica
symmetric breaking performs and the solution
subspace has many different macroscopic states. Besides,
the whole solution space still has great complexity in finding different \emph{proper selections}, i.e., different solution sub-spaces.
Therefore, combing the complicated organization of the selections and the replica symmetric breaking in each
selection, two stages of complexity arise for the Vertex-Cover solution space and we think it should be the essence of the further-step replica symmetry breaking phenomena \cite{wolf} in Vertex-Cover.

Then, some discussions on vertex status environment will be provided in the following.
By the above analysis on the unfrozen vertices connectivity, we can have some facts on the organizations
of the detailed organizations in the steady state of the reduced solution graph.

$\diamond$ One vertex belongs to positive backbones if and only if
all its neighbors are negative backbones. This is always true under any condition.

$\diamond$ One vertex is unfrozen if and only if it belongs to a mutual-determination,
and a mutual-determination can exist only when
all the neighbors of its two vertexes are unfrozen or negatively frozen, without
positive backbones. It is a necessary condition for the mutual-determination environment but
not a sufficient one.

$\diamond$ One vertex belongs to negative backbones only when
at least one of its neighbors is positively frozen. It is a sufficient and necessary condition only when
there are no \emph{odd cycle breaking} operations \cite{space-vc} performed during
the process of obtaining the steady reduced solution graph.

\begin{figure}[!t]
\centering
\includegraphics[width=5in]{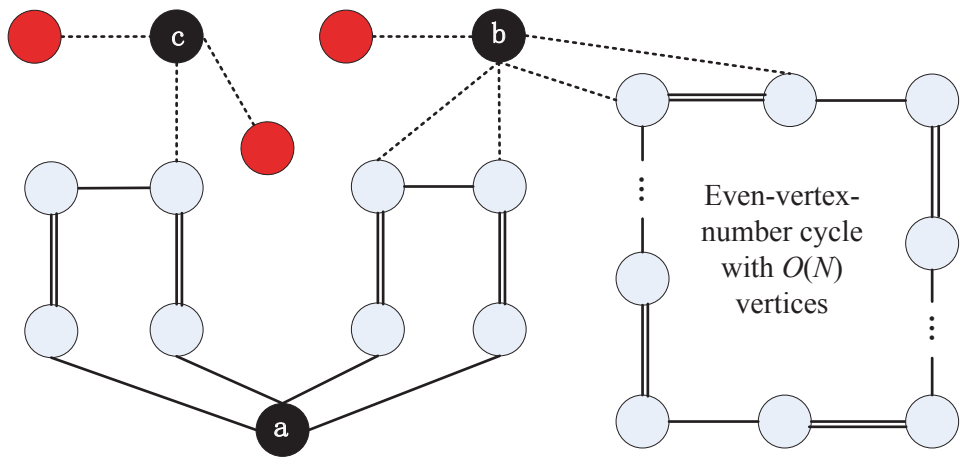}
\caption{The exact solution space expression of an example
for a schematic view of complicated environment of possible organizations. For the negative backbones $a$, $b$ and $c$,
there are 0, 1 and 2 positively frozen neighbors. And, for the even-vertex-number cycle on the right part of the graph, long-range
correlation works and produces two different macroscopic states.}
\label{fig1}
\end{figure}

For the sufficiency of the second fact,
in Fig.3, vertex $b$ and its red neighbor both have their neighbors to be
unfrozen, but they do not form a mutual-determination but backbones.
For the third fact when \emph{odd cycle breaking} operations \cite{space-vc} perform,
the negative backbone can have no positively frozen neighbors,
one positively frozen neighbor or more, and the complicated environment of negative backbones are shown in Fig.3.
In \cite{long-range-zhou}, two conjectures are assumed: (I) if a vertex is positively frozen in one state,
it is positively frozen in all states, and (II) a vertex is
negatively frozen in all states only if it is adjacent to two
or more positively frozen vertices. However, by our analysis above, there are something inaccurate
for these two conjectures: For (I), shown in Fig.3, as the existence of the even-vertex-number cycle with length
$O(N)$, the solution space splits into two different clusters which have different vertices' values on the even-vertex-number cycle,
that is, positively frozen vertices on the even-vertex-number cycle in one cluster must be negatively frozen ones in another.
For (II), it is easy to see that it is not always correct, which has been discussed in the third fact.
However, as the example here is a constructed one and the above
analysis is for all the cases, it is thought that these conjectures may work well with high probability for random Vertex-cover instances.

\section{Solution number counting on Vertex-Cover}

Next, we will concern on the solution number counting of Vertex-Cover, provide an algorithm
dealing with it, and then
give the exactness and efficiency analysis for the algorithm.

\subsection{Solution number counting algorithm for Vertex-Cover}

Till now, statistical mechanics has provided calculation methods under the assumptions of replica
symmetry or replica symmetric breaking for the entropy of Vertex-Cover solutions \cite{vc-bp}.
But, these methods for the solutions' entropy mainly focus on statistical analysis and always lack rigorous theoretical supports.
In this subsection, we aim to calculate the solution numbers based on the reduced solution graph in a rigorous way. Thus,
we only consider the obtained reduce solution graph which can exactly
express the whole solution space by the MBEA algorithm \cite{space-vc}.
Evidently, this cannot always be guaranteed when the replica symmetric breaking works.

For a random graph $G$ with $n$ vertices, its unfrozen subgraph is also of local tree-like structure and can only
have cycles of at least $O(log(n))$ scale. For $c<e$, there is no giant unfrozen component with high probability,
so almost no cycles on the unfrozen subgraph can exist. When the unfrozen subgraph is a tree,
the accurate solution number of Vertex-cover can be achieved using the cavity method.
By adding vertices from the leaves to the root of the tree hierarchically, a sub-tree is obtained
after adding a new vertex $i$ in each step. We can define $S(i)$ as the solution number of the current sub-tree,
and $S^+(i),S^-(i)$ are the solution numbers of the current sub-tree when vertex $i$ takes $+1$ (uncovered) or $-1$ (covered). Then
by the right figure of Fig.1 where vertex $l$ is a high level one over vertex $i$, we have in case (a)
\begin{equation}
S^+(i)=\prod_{j=1}^{h}S^-(k_j), \\S^-(i)=\prod_{j=1}^{h}S(k_j),
\end{equation}
and in case (b)
\begin{equation}
S^+(i)=S^-(k)\cdot\prod_{j=1}^{h}S^-(k_j), \\S^-(i)=S^+(k)\cdot\prod_{j=1}^{h}S(k_j),
\end{equation}
where $S(k_j)=S^+(k_j)+S^-(k_j)$. Iterating the formula from the leaves to the root on the unfrozen subgraph,
the total number of solutions can be obtained as $S(root)$. When the unfrozen subgraph is a forest $T$ of trees $T_1, \cdots, T_s$,
equations (2-3) also work for each connected component, and the total number of solutions can be expressed as
\begin{equation}
     S(T)=\prod_{k=1}^s  S(T_k),
\end{equation}
where $S(T_k)$ is the solution number of $T_k$ and $S(T)$ is the total solution number of forest $T$. Here, the
whole time consumption of the algorithm is $O(n)$.

If the unfrozen subgraph has cycles, the above method will not be an accurate one, and a modified kind
of exhaustive method should be used. For those with fewer cycles on the unfrozen subgraph which
can become a forest or a tree after deleting $k$ (no more than $O(log n)$) vertices, simply having an
exhaustion on the status of these vertices will produce $2^k$ subproblems with tree structure, and the whole
time consumption is polynomial. In our solution number calculation, for the obtained unfrozen subgraph by MBEA algorithm,
the cycles will be broken by considering the unfrozen vertices with

\noindent
\begin{tabular}[c]{p{480pt}l}  \textbf{Vertex-Cover Solution Number Counting
Algorithm}\\
\hline \hline \textbf{INPUT:}  \texttt{Unfrozen graph $G_u$ } \\
\textbf{OUTPUT:} \texttt{Solution Number $S_{vc}(G)$}\\
\hline
\textbf{\emph{Tree-Counting}} \ (\texttt{Tree} $T$)\\
\textbf{begin}\\
\ \ \ \ \ {\texttt{define leaf-removal order from leaves to the root of} $T$}\\
\ \ \ \ \ {\texttt{set the states of each leave $S^+(leaf)=S^-(leaf)=1$}}\\
\ \ \ \ \ \textbf{for} \ (\texttt{each non-leaf vertex} $i$ \texttt{on} $T$ \texttt{by the order})\\
\ \ \ \ \ \ \ \ \ \ \ \textbf{if} ({\texttt{vertex $i$ has a double-edge child $k$ and $h$ single-edge children}})\\
\ \ \ \ \ \ \ \ \ \ \ \ \ \ \ \ \ {$S^+(i)=S^-(k)\cdot\prod_{j=1}^{h}S^-(k_j)$}\\
\ \ \ \ \ \ \ \ \ \ \ \ \ \ \ \ \ {$S^-(i)=S^+(k)\cdot\prod_{j=1}^{h}S(k_j)$}\\
\ \ \ \ \ \ \ \ \ \ \ \textbf{else} ({\texttt{vertex $i$ has $h$ single-edge children}})\\
\ \ \ \ \ \ \ \ \ \ \ \ \ \ \ \ \ {$S^+(i)=\prod_{j=1}^{h}S^-(k_j)$}\\
\ \ \ \ \ \ \ \ \ \ \ \ \ \ \ \ \ {$S^-(i)=\prod_{j=1}^{h}S(k_j)$}\\
\ \ \ \ \ \ \ \ \ \ \ {$S(k_j)=S^+(k_j)+S^-(k_j)$}\\
\ \ \ \ \ \textbf{return($S(root)$)}\\
\textbf{end}\\
\textbf{\emph{Forest-Counting}} \ (\texttt{Forest} $F_T$)\\
\textbf{begin}\\
\ \ \ \ \ $S(F_T)=1$\\
\ \ \ \ \ \textbf{for} \ (\texttt{tree $T_i$ in forest $T$} )\\
\ \ \ \ \ \ \ \ \ \ \ {\texttt{$S(T_i)=$\textbf{\emph{Tree-Counting}($T_i$)}}}\\
\ \ \ \ \ \ \ \ \ \ \ {$S(F_T)=S(F_T)*S(T_i)$}\\
\ \ \ \ \ \textbf{return($S(F_T)$)}\\
\textbf{end}\\
\textbf{\emph{main}} \ (\texttt{unfrozen graph} $G_u$)\\
\textbf{begin}\\
$S_{vc}(G_u)=0$\\
\textbf{if} \ ($G_u$ \texttt{has cycles})\\
\ \ \ \ \ \textbf{do}\\
\ \ \ \ \ \ \ \ \ \ \ \textbf{for} \ (\texttt{each vertex} $i$ \texttt{on} $G_u$)\\
\ \ \ \ \ \ \ \ \ \ \ \ \ \ \ \ \ {\texttt{calculate influence range}\ $I_j=max_i \ I_i$ }\\
\ \ \ \ \ \ \ \ \ \ \ {\texttt{do exhaustion on vertices status with vertex} $j$}\\
\ \ \ \ \ \ \ \ \ \ \ {\texttt{renew} $G_u$ \texttt{by deleting} $j$ \texttt{and vertices in its influence range}}\\
\ \ \ \ \ \textbf{while}\ ($G_u$ \texttt{has cycles}) \\
\ \ \ \ \ \textbf{for}\ (\texttt{each exhaustion on above vertices $j$s }) \\
\ \ \ \ \ \ \ \ \ \ \ {\texttt{$S(G_u)$=\textbf{\emph{Forest-Counting}($G_u$)}}}\\
\ \ \ \ \ \ \ \ \ \ \ {$S_{vc}(G_u)=S_{vc}(G_u)+S(G_u)$}\\
\ \ \ \ \ \textbf{return($S_{vc}(G_u)$)}\\
\textbf{else} \\
\ \ \ \ \ {\texttt{$S_{vc}(G_u)$=\textbf{\emph{Forest-Counting}($G_u$)}}}\\
\ \ \ \ \ \textbf{return($S_{vc}(G_u)$)}\\
\textbf{end}\\
\}
\end{tabular}
\noindent

\noindent greatest influence.
Under the information propagation on the unfrozen subgraph, the number of
vertices $I_i^+$ affected by
vertex $i$ uncovered and $I_i^-$ by $i$ covered can be easily determined. When breaking the cycles,
we do the exhaustion on the status of the vertices with maximum influence ranges $I_i=I_i^++I_i^-$,
which is shown in Fig.1 as the top level mutual-determination vertices. The
Vertex-Cover Solution Number Counting Algorithm is shown as follows and an example is given in Fig.4.

\subsection{Complexity and performance analysis for the algorithm}
\begin{figure}[!t]
\centering
\includegraphics[width=5.5in]{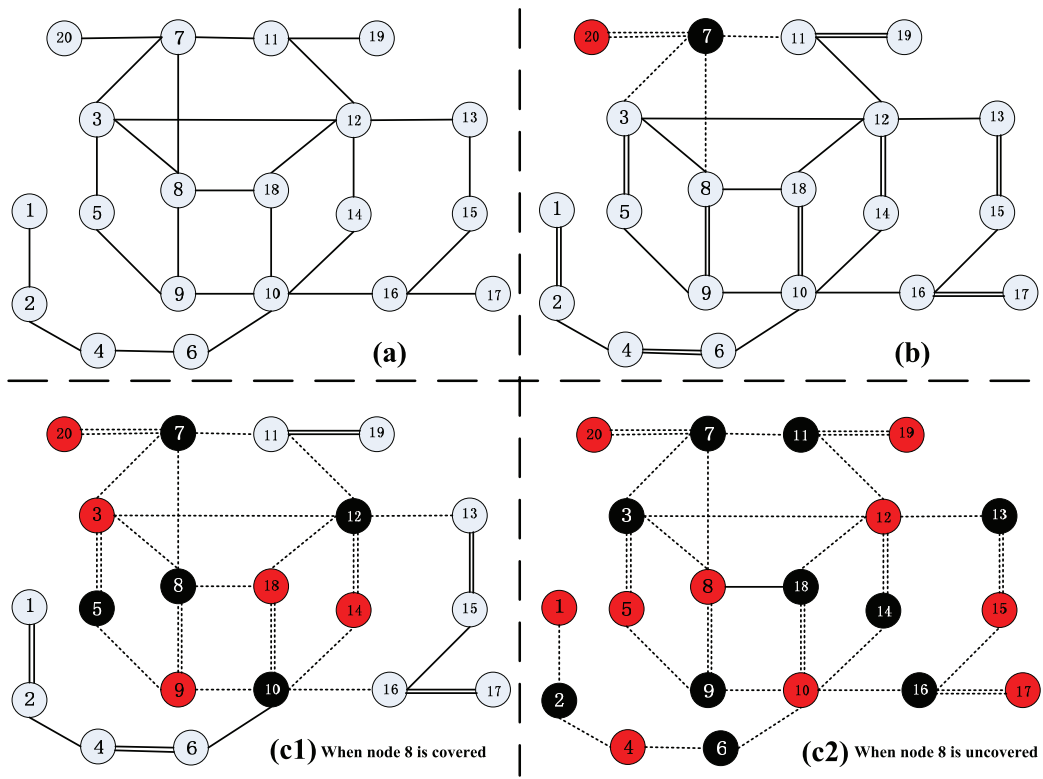}
\caption{An example for counting the solution number of Vertex-Cover.
Subgraph (a) is the original graph and subgraph (b) is the corresponding
reduced solution graph. Subgraph (c1) is the case when vertex 8 is covered, and the corresponding
unfrozen subgraph has three connected components, in which there are 2*3*3=18 solutions.
Subgraph (c2) is the case when vertex 8 is uncovered, the corresponding
unfrozen subgraph is null and all vertices are frozen, so there is only one solution in this case.
Thus, for this example, the total number of solutions is 19.}
\label{fig1}
\end{figure}

In the frame of our leaf-removal levels in Fig.1, for each exhaustive assignment of the top-level mutual-determination vertices,
a great number of unfrozen vertices (nearly a half in
probability) should be fixed by the requirement of Vertex-Cover and mutual-determination
relations.
Supposing there are $n_k$ mutual-determinations in the top level $k$, the
exhaustive assignments number of this level is $2^{n_k}$, each of which will cause about a half unfrozen vertices to be frozen
and the unfrozen subgraph greatly contracted. Thus, defining $C(n)$ as the complexity of counting
the solution number of an unfrozen subgraph with $n$ vertices, we have $C(n)\doteq 2^{n_k}C(n/2)$ after exhausting
the top level $k$. Then, the original problem is reduced to sub-problems with sizes about $n/2$, and newly produced top levels
in each sub-problem can be located with a new round of exhaustion.

As a result, if the number of mutual-determinations in each stage of exhaustive top levels is of at most $L=O(1)$,
the total complexity $C(n)$ will be polynomial ($O(n^L)$ in the worst case) by recursive
solving. If the number of mutual-determinations in each stage of exhaustive top levels is of at most $O(log(n))$,
the total complexity $C(n)$ will be super-polynomial and sub-exponential ($O([log(n)]^{log(n)})$ in the worst case) by recursive
solving; but if only $O(1)$ exhaustive top levels in the stages have $O(log(n))$ mutual-determinations and
the others have $O(1)$ mutual-determinations,
the total complexity is still polynomial.
If there exists at least one exhaustive top level with $O(n)$ mutual-determinations,
the complexity by this strategy will be exponential.

Furthermore, the above strategy can be revised to perform more efficiently.
First, if the sub-problems after some exhaustion stages are with the form of a tree
or a forest, the Forest Counting Algorithm in our algorithm should be performed;
if the sub-problems are with the form of
unconnected subgraphs, they can be handled with different unfrozen components separately
and the complexity will greatly decrease.
Besides, as there are many even-vertex-number cycles (cycles with $2k$ unfrozen vertices
and $k$ mutual-determinations, $k=2,\cdots$) on the unfrozen subgraph,
all the vertices on such a cycle can be viewed as an equivalent class, which means that the fixation of
each vertex will cause fully fixation of all the other vertices on this class.
Thus, treating vertices on one such even-vertex-number cycle
as one unfrozen vertex can greatly reduce the size of the unfrozen subgraph.
At last, the above exhaustive levels are actually the top levels in each (sub-)problem,
however, for each specific instance, it is not necessary to only choose the top levels of each (sub-)problem as exhaustive levels,
and the exhaustion on some next-top levels would produce the similar effects. Therefore, the strategy for counting the solutions
can be modified by choosing exhaustive levels with relatively fewer mutual-determinations
nearby the top levels.

For a random graph $G$, its unfrozen subgraph can be handled by the above
strategies, and mean entropy density $s(c)=logS(G,c)/n$ is calculated in Fig.5.
In Fig.5, our algorithmic results (the red squares) fit very well with
the results of 1RSB cavity method and the simulations by \cite{martin1,vc-entropy}.
Especially in the neighborhood of the phase transition point $c=e$,
our algorithmic results approach the simulation results much better than those by RS and 1-RSB.
For the NP-completeness of Vertex-Cover problem, the performance of the MBEA algorithm for obtaining the
reduced solution graph cannot be completely exact when $c>e$ and the replica symmetric breaking works.
Though it is a quite small difference between the obtained MEBA-algorithm coverage and the true minimal coverage of a given graph
even when the average degree $c$ is relatively large \cite{space-vc}, this small difference on the minimum energy
can lead to meta-stable states of Vertex-Cover and an increase of entropy by our proposed solution-number counting algorithm.
Thus, in the implement of the above Vertex-Cover Solution Number Counting
Algorithm, we make verification of the MEBA-algorithm results, and perform our algorithm
only on the exact instances in which the reduced solution graphs have exact minimal coverage.
During the verification, we find that the MBEA algorithm can work very well when $c<4$ and it provides
almost the exact minimal coverage for all instances, but its exactness decreases as the average degree $c$ increases over $4$.
However, our proposed Vertex-Cover Solution Number Counting
Algorithm always has good performance given an exact reduced solution graph.
It should be quite interesting to study how to improve the MEBA algorithm to make it perform much better,
which will be focused in our future research.

\begin{figure}[!t]
\centering
\includegraphics[width=4in]{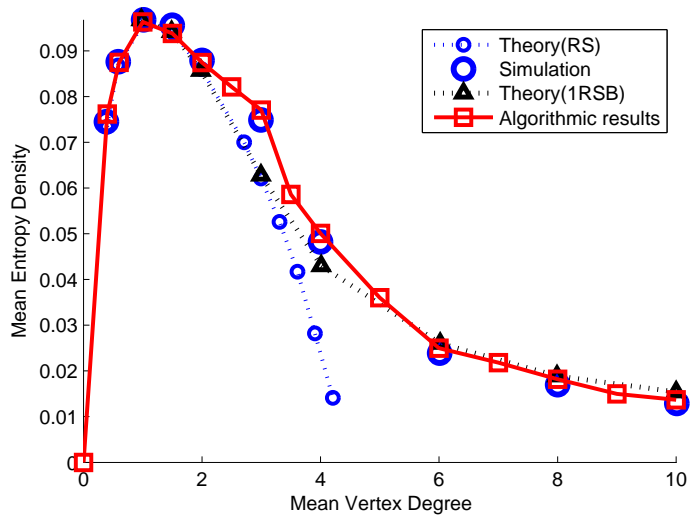}
\caption{The mean entropy density for random graphs with average degree $0<c<10$. The algorithmic results are
achieved by 10000 instances with $n=1000$ vertices, which fit very well with those of simulations and the 1RSB cavity method.
The replica symmetric results are also provided as a comparison \cite{vc-entropy}.  }
\label{fig1}
\end{figure}

\section{Analysis on the characteristics of solution number statistics}

Based on the above analysis of solution organizations and solution numbers,
some further analysis on solution number statistics, such as
solution number variation and fluctuation and marginal probabilities, are investigated
in this section.

\subsection{Solution number variation and fluctuation}

We first consider the solution number variation by an increasing process of double edges.
When a new double edge is added to the unfrozen subgraph,
its two ends will connect other mutual-determinations. It is easy to see that if the new double edge connects
different connected components on the unfrozen subgraph (a) or it enhances the connection in one connected component (b),
the number of solutions will decrease; only when the new double edge is an isolated one (c) or if it has only one-end connections
with other mutual-determinations (d), the number of solutions can increase. In case (a), if a new double edge connects two
unfrozen connected components $C_1, C_2$ whose solution numbers are $n_1, n_2$, then the total
solution number will decrease by a factor from $n_1\times n_2$ to $n_1+n_2$; if more unfrozen components are connected,
the total solution number will decrease more sharply.
In case (b), a new added double edge in an unfrozen connected component whose solution number is originally $n$,
can lead to a decrease of $0$ to $n-2$ on the solution number according to the structure of the component
and the position of the new added double edge.
In case (c), the total solution number increases by a factor $2$.
In case (d), the increase of the total solution number is at least $1$. By the results in Fig.5, we can see
that the maximum of the entropy is achieved at about $c=1$ for random graph. Thus,
before $c=1$, cases (c) and (d) dominate the solution number variation and lead to its increase;
after $c=1$, cases (a) and (b) dominate the solution number variation and lead to its decrease.
This point accords with that of random graph percolation \cite{rg}, by which only small components connectivity
increases before $c=1$, and it may be the reason why case (c) and (d) can dominate the solution number variation.

Then, we analyze the solution number fluctuation by the influence range of one unfrozen vertex being uncovered.
Given a double edge on the unfrozen subgraph and considering one of its end, it being uncovered will cause
other $s$ unfrozen vertices in its \emph{influence range} to be fixed with probability
\begin{equation}
P(s)=\sum_{k=1}^{\infty}F_r(k)\prod_{j=1}^{k-1}P(s_j)\delta(\sum_{j=1}^{k-1}s_j=s-1).
\end{equation}
Here, when the current unfrozen vertex $i$ is chosen to be uncovered, its mutual-determination neighbor
must be covered and the influence in this direction terminates; the other unfrozen neighbors of $i$ should also be
covered and their correspondingly mutual-determination neighbors should be uncovered; and then influence
can be propagated by the new uncovered vertices similar as vertex $i$.

By the results in \cite{long-range-zhou}, a structure named long-range frustration is well studied,
which reveals a set of vertices having great influence on others.
As the definition of long-range frustration vertices, their being uncovered will cause a large number ($O(N)$) of vertices
to be frozen. By calculating $R=1-\sum_{s=0}^{\infty}P(s)$, we can obtain the ratio of
vertices with great influence ($O(N)$ vertices). By the organization of the unfrozen subgraph, one leaf on it being covered may also cause large influence
on the graph. In the leaf-removal \emph{core} of the unfrozen graph, two vertices of one mutual-determination may both
have $O(N)$ influence ranges, they possess great impact on the solution space correlations and solution number counting,
and correspond to the vertices in the top levels of Fig.1. In the literature
of mutual-determination \cite{space-vc}, the existence of the long-range frustration vertices can be
supported by the unfrozen vertices of top levels or high centrality \cite{centrality} in the leaf-removal core.
Thus, the change of one such vertex's status may greatly reduce the unfrozen subgraph sizes and
correspondingly the solution numbers.

As the existence of top-level unfrozen vertices with large influence ranges, the solution number will fluctuate by increasing $c$.
When a new edge is added to the original graph, it may affect some such top-level vertices to be frozen, which can
produce great influence to the unfrozen subgraphs. Then, the unfrozen subgraph and solution space sharply contract and collapse,
leading to great reduction of the solution numbers. However, these frozen top-level unfrozen vertices with their influence
ranges are easily to recover their unfrozen states in the subsequent process of adding edges by the \emph{Releasing Operations}
in the MBEA algorithm \cite{space-vc}. Thus, there may exist large fluctuations on the solution number when a single edge is added, but
these fluctuations can be neglected by average in a time period within $O(N)$ steps.
Therefore, the number of unfrozen vertices and mutual-determinations increases by average degree $c$
in a macroscopic view, which supports the statistical results in \cite{vc-bp,martin1,martin2,wolf,long-range-zhou}.

\subsection{Calculation of marginal probability on solution space}

Furthermore, we will use a similar analysis as the Vertex-Cover Solution Number Counting
Algorithm to calculate the marginal probability distribution of Vertex-cover,
which is an important statistical quantity of partition functions \cite{partif} in statistical mechanics.
When the unfrozen subgraph is with of few cycles, the marginal probability $P(x_r=+1)=P^+_r$ of vertex $r$
can be determined by the following steps:

\emph{Step 1}: Choose vertex $r$ as a root of its connected component on the unfrozen subgraph,
and initialize the marginal probability of the leaves by a probability $P^+(leaf)=0.5$.
If root $r$ is also a leaf, do not give it any initialized value.

\emph{Step 2}: Iterate the following formulas similarly as those in equations (2-3) from the leaves to the root,
\begin{equation}
    P^+(i)=\prod_{j=1}^{h}(1-P^+(k_j)), \ \ \ \ \ or \ \ \ \ \
    P^+(i)=(1-P^+(k))\cdot\prod_{j=1}^{h}(1-P^+(k_j)).
\end{equation}
This equation also uses the notations in the right figure of Fig.1, which has a correspondence
with cases (a-b) respectively.

\emph{Step 3}: Finally, the local environment of vertex $r$ can be fixed, and
$P^+(r)$ is obtained which is just the marginal probabilities $P^+_r$ and $P^-_r=1-P^+_r$.
Note that none of the $P^+(i)$s except $P^+(r)$ in this process are the marginal probability $P^+_i$ of the
whole graph but only some intermediate variables.

After the above steps, we can obtain the marginal probability of different vertices by choosing
different roots, and all the marginal probabilities can be obtained in about $O(n^2)$ steps.
Besides, we can directly use the results of the Vertex-Cover Solution Number Counting
Algorithm, calculate the total solution number and the solution number after fixing the considered vertex,
and then the marginal probability is achieved.
The marginal calculation can be (almost) accurate on the (almost) tree structures,
the validity conditions of which are the same as the Vertex-Cover Solution Number Counting
Algorithm.

By the above method, marginal probabilities of Vertex-Cover on random graph are given in the left figure of Fig.4,
and the vertices can be nearly classified into 3 classes: positive backbones $P^-_r=0$, nearly negative 
backbones $P^-_r\doteq 1$, and \emph{almost completely unfrozen vertices} $P^-_r\doteq 0.5$. The other unfrozen vertices
occupy a small proportion and the ratio of negative backbones increases
with the average connectivity. Thus, in Vertex-cover, the status of vertices possesses strong polarization phenomenon,
and it can be recognized as the reason why survey propagation algorithm can have
good performance in finding the minimal vertex-covers \cite{martin1,martin2,vc-entropy}.

\begin{figure}[!t]
\centering
\includegraphics[width=6in]{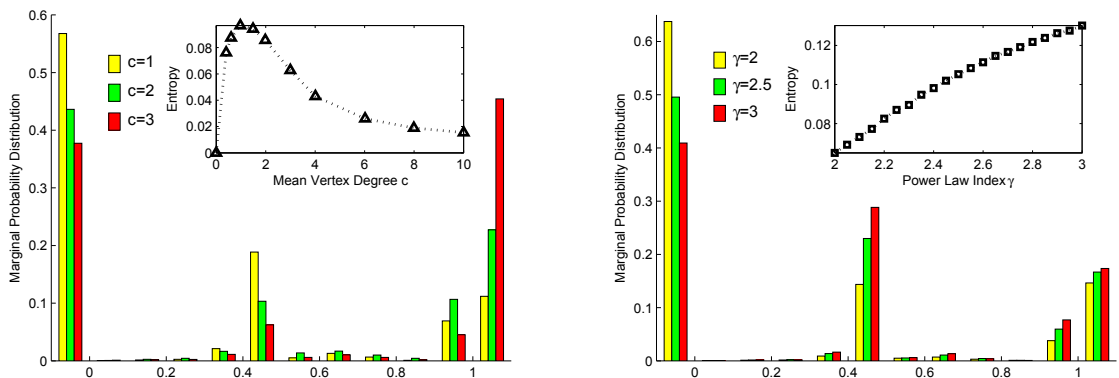}
\caption{The Vertex-Cover marginal probability distribution for random graphs with $c=1,2,3$ and scale-free graphs
with $\gamma=2,2.5,3$. In the insets, mean entropy densities (1RSB results for random graph and our algorithmic results
for scale-free graph) are shown separately.
The horizontal axis
gives the probability of a randomly chosen vertex being covered $P^-_r$, and the vertical axis
gives the ratio of vertices having the same probability. The horizontal axis
is uniformly divided into 10 intervals, with the left and right data for positive and negative backbones.
The results are achieved by 10000 instances with $n=1000$ vertices.}
\label{fig1}
\end{figure}

At last, to see the effect of our algorithm on graphs with different structures,
we perform it on Vertex-Cover of scale-free graphs \cite{martin3}. For a randomly generated scale-free
graph with degree distribution $P(k)\sim k^{-\gamma}$, the
solution number counting
and marginal probability calculations are done separately,
which are shown in the right figure of Fig.4 with $2 \leq \gamma \leq3$. There are
similar phenomena with marginal probabilities on random graphs, but the ratio of almost completely unfrozen vertices monotonically
increases with $\gamma$.



\section{Conclusion and discussion}
Some underlying organizations such as degree distributions of
unfrozen vertices and component sizes are investigated in this paper.
Based on the organization of unfrozen subgraph, our Vertex-Cover Solution Number Counting
Algorithm
can accurately calculate the solution number of Vertex-Cover in polynomial time
when the unfrozen subgraph is accurate and has relatively few cycles.
The algorithm can give solution number for instances, does not rely on the graph structures heavily
and works better in the region of the phase transition point $c=e$ on random graphs.
Besides, by the algorithm, variation and fluctuation after adding an individual edge are studied on solution numbers, and
marginal probabilities on random graph and scale-free graph are provided.

Further research
on the scaling window \cite{compo-size} of $c=e$
will benefit for the understanding of the essence of the phase transition.
Besides, as the proposed algorithm is mainly based on the obtained unfrozen subgraph,
it is still important work to improve the performance of MBEA algorithm and make the reduced
solution graph more accurate, which needs
more technical tackling and optimized strategies.
More strategies should be investigated to manage the coverage approximations
and obtain more accurate solution counting results, which can lead to better cognition
on the formation and statistics of the hardness of NP-complete and \#P-complete problems.

\section*{Acknowledgement}
This work is supported by the Fundamental Research Funds
for the Central Universities and the Natural Science Foundation
of China (Grant No. 11201019).
\\

\bibliography{basename of .bib file}

\end{document}